\documentclass[10pt, conference, letterpaper]{IEEEtran}

\usepackage{amsthm,amsmath,amssymb,esint,color,graphicx,overpic,bbm,cite,wrapfig,acronym,subcaption,latexsym,paralist,xspace,array,multirow,url,framed, caption}
\usepackage{algorithm, tikz}
\usepackage[noend]{algpseudocode}

\usepackage[font=small,labelfont=bf]{caption}

\newtheorem{example}{Example}

\DeclareMathOperator*{\maximize}{maximize}
\begin{document}

\title{Learn2MAC: Online Learning Multiple Access \\ for URLLC Applications}
\author{
	\IEEEauthorblockN{Apostolos Destounis, Dimitrios Tsilimantos, M\'erouane Debbah and Georgios S. Paschos
		\\}
	\IEEEauthorblockA{Mathematical and Algorithmic Sciences Lab, France Research Center, Huawei Technologies Co. Ltd. \\
		Arcs de Seine, 20, quai du Point du Jour, $92100$, Boulogne-Billancourt, France\\
		email: firstname.lastname@huawei.com
	}
}
\maketitle

\maketitle
\begin{abstract}
This paper addresses a fundamental limitation of previous random access protocols, their lack of  latency performance guarantees. We consider $K$ IoT transmitters competing for uplink resources and we design a fully distributed protocol for deciding how they access the medium. Specifically, each transmitter restricts decisions to a locally-generated dictionary of transmission patterns. At the beginning of a frame, pattern $i$ is chosen with probability $p^i$, and an online exponentiated gradient algorithm is used to adjust this probability distribution. The performance of the proposed scheme is showcased in simulations, where it is compared with a basline random access protocol. Simulation results show that (a) the proposed scheme achieves good latent throughput performance and low energy consumption, while (b) it outperforms by a big margin random transmissions.
\end{abstract}

\section{Introduction}

In the upcoming \emph{Internet of Things} (IoT) an immense number of devices will be connected to each cellular station--forecasts predict 1 million devices per station \cite{report}. IoT connectivity  is primarily aimed at establishing central authentication, security, and management of those devices. However, fine-tuned coordination functionalities (transmit power selection, transmission scheduling, code assignment, etc) are considered very expensive to be handled centrally, since the cellular station would need to collect a bulky state information for each device and solve large-scale optimization problems. For these reasons, it is anticipated that IoT communications will rely on \emph{uncoordinated access}, i.e., a channel will be dedicated to IoT access and each IoT transmitter will decide individually which transmission pattern to use. Here, we study the use of Online Learning methods for transmission pattern selection.

\begin{figure}
  \centering
     \includegraphics[width=0.45\textwidth]{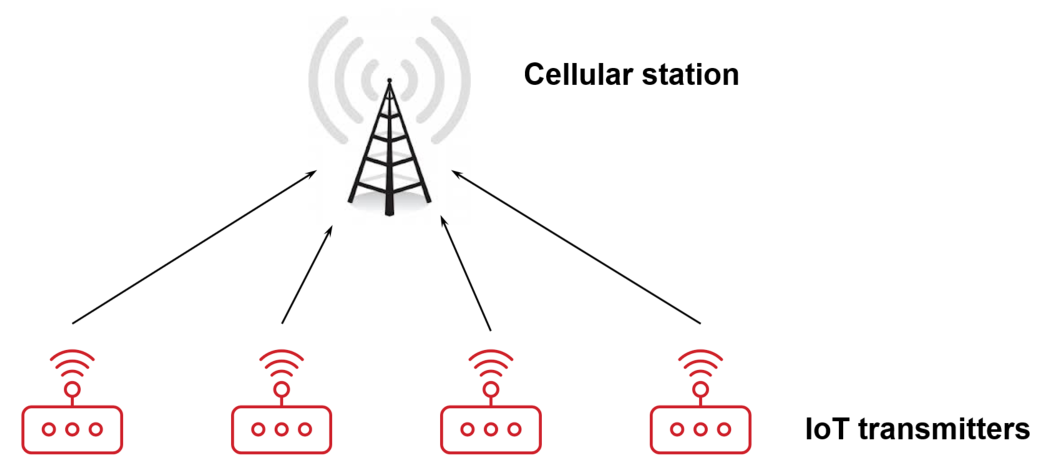}
  \caption{IoT transmitters share a common wireless medium in an uncoordinated manner.}
  \label{fig:SystemModel}
\end{figure}

We consider $K$  transmitters scattered in a geographical area, all wanting to transmit to the cellular station (e.g. a common sink), as shown in \figurename{ \ref{fig:SystemModel}}. We further assume that a) for reasons of overhead reduction, there is no coordination between a transmitter and the cellular station, and b) for reasons of security there is no coordination among different transmitters. Each transmitter must decide on its own when and how to transmit.

\subsection{Random access protocols}

Traditional protocols that can operate in this setting are based on random access. Historically, pure ALOHA  was the first such protocol, where a user transmits with a probability $p$ \cite{pure_aloha}. This was later extended to slotted-ALOHA \cite{aloha_slotted}, which used synchronization to double user throughput. 
A more mature random access protocol is the \emph{Carrier Sense Multiple Access} (CSMA), where the transmitter  checks whether the medium is idle before sending. Also, in the enhanced version with collision avoidance (CSMA/CA) the transmitter ``backs-off'' (selects a smaller probability of access) every time there is a collision, while also uses ready-to-transmit (RTS) and clear-to-transmit (CTS) signals to reduce the impact of a collision on throughput  \cite{bianchi00}. 

Random access protocols suffer from  collisions and idle time, and therefore they achieve lower throughput than the maximum possible. 
In an effort to improve the throughput achievable by uncoordinated access, many exciting algorithmic ideas have been proposed. For example, Q-CSMA \cite{NiSrikant}  is a protocol where the transmitters avoid collisions by finding efficient schedules in a distributed manner (see also \cite{JiangWalrand}). Although Q-CSMA is shown to asymptotically achieve 100$\%$ throughput (maximum possible),  it suffers from large delays. 
Another interesting direction is the idea of successive cancellation and replica transmission \cite{SuccesiveCanc}. In this enhanced random access protocol, each transmitter sends multiple replicas of the same packet within a frame. Normally, a large number of collisions occur, but with the assumption that the Signal-to-Interference-plus-Noise (SINR) levels of transmitters are relatively different, the receiver can decode the strongest one, subtract it from the next, etc, and eventually decode correctly all signals. This protocol achieves high throughput, but at the cost of excessive energy usage, which is a concern in IoT applications.

\subsection{Communication requirements for IoT}

We list our requirements for IoT communications.

\subsubsection{URLLC}

The \emph{Ultra Reliable Low Latency Communications} (URLLC) class is a popular 5G definition for communications of high fidelity, seen as an enabler for remote control of vehicles, and other demanding applications. In URLLC, a given amount of bits must be received before a strict deadline (in periods) with a very high probability (often 0.99999). This reliability guarantee is extremely important in automation and remote control, as well as in applications where freshness of information is essential, and the operation of some IoT applications will rely on such guarantees. For this reason, we depart from  pure throughput considerations, and we define below the \emph{latent throughput}, which suffices to meet URLLC requirements. 

Time is split in frames and within each frame there are $N$ slots. A frame is then called ``successful for transmitter $k$'' if it contains $L$ or more successful transmissions of transmitter $k$.
Successful transmissions in previous frames do not count towards the success criterion of the current frame. The latent (URLLC) throughput is the empirical frequency of successful frames. 
We note that no existing random access protocol provides latent throughput guarantees, as all of them are designed for maximizing pure throughput which is different from latent throughput. 
For example, $L-1$ successful transmissions within a frame provide $\frac{L-1}{N}$ pure throughput, but amount to $0$ latent throughput.
More generally, latent throughput optimization is a difficult problem even with centralized coordination \cite{Apostolos2}, and has strong ties to the theory of Markov Decision Processes \cite{Apostolos1}.

\subsubsection{Energy consumption}

Since the majority of IoT devices will work on batteries, energy consumption must be minimized. In this work we assume that energy is proportional to the number of transmissions. 

\subsection{Our contribution}

In this paper we propose a protocol for uncoordinated medium access, which is based on the theory of Online Learning \cite{Shalev}. First, we restrict our transmitter to choose transmission patterns in the beginning of the frame, and in particular, we further restrict its options to a randomized dictionary of patterns. During operation, the transmitter first chooses a pattern from the dictionary at random, and then implements the pattern within the frame. The learning operation amounts to progressively adjust the probability distribution of pattern selection using an online exponentiated gradient descent algorithm. Our simulations show that the resulting Learn2MAC scheme:
\begin{itemize}
\item Achieves high URLLC throughput and low energy consumption, when faced against (i) TDMA interference, or (ii) Random access interference.
\item Multiple Learn2MAC users can outperform, in terms of latent throughput, the ALOHA users by as much as 100\%.
\end{itemize}

\section{Problem formulation}

\subsection{System model and assumptions}
There are $K$ transmitters sharing the uplink of our system. 
Time is split in frames of $N$ slots. At the beginning of frame $t$,  transmitter $k$ decides a \emph{pattern of transmissions} to be used within the frame; we denote this decision with $x_k(t)\in \{0,1\}^N$, where $x_{k,n}(t)=1$ indicates transmission in slot $n$, and $x_{k,n}(t)=0$ indicates idling.
Therefore, at each frame a transmitter chooses its pattern as a binary vector of length $N$ from the set $\mathcal{X}=\{0,1\}^N.$

Our pattern selection setting is very general, as the next example suggests.
\begin{example}[ALOHA]
Consider $N=2$, where all possible transmission patterns are $\mathcal{X}=\{(0,0),(1,0),(0,1),(1,1)\}$. A simple protocol could be: ``choose one pattern at random with probability 1/4 independently of past events''. Incidentally, this corresponds to a slotted-ALOHA with $p=1/2$.
\end{example}

We make the following assumptions about our system. 

\begin{itemize}
\item[\textbf{(A.1)}] If two or more transmitters have selected to transmit at the same slot, we have a collision and all transmitted information in this slot is lost.\footnote{In this paper we study the ``hard interference'' scenario for simplicity. We mention, however, that our work can be extended to other interference models.}

\item[\textbf{(A.2)}] At the end of frame $t$, the cellular station provides  feedback  information about the occupancy of each slot (idle/success/collision) to all transmitters.  
\end{itemize}

We reserve $x_k(t)$ to denote the pattern selected by user $k$ in frame $t$, and $\pi$ to index patterns in the set $\mathcal{X}$.
Because of \textbf{(A.1)}, 
a pattern $\pi \in \mathcal{X}$ produces \emph{a successful transmission for user $k$} in slot $n$ (an event denoted with $s_{k,n}(\pi)=1$)  only if $\pi_{\ell,n}=0,~~\forall \ell\neq k$. Equivalently, we write:
\begin{equation}\label{eq:success}
s_{k,n}(\pi)= \pi_{k,n}\prod_{\ell\neq k}(1-\pi_{\ell,n}), ~~\forall \pi\in\mathcal{X}.
\end{equation}

\subsection{Performance metrics}

Our protocol design is driven by certain objectives, which are used to form the utility function of each transmitter.  

\textbf{URLLC throughput.}
In frame $t$ a pattern $\pi\in\mathcal{X}$ is called \emph{successful}, denoted with $R_t(\pi)=1$, if it contains at least $L$ successful transmissions.\footnote{$L$ in this case is an application-specific parameter that captures the amount of successful transmissions required within a frame in order for user $k$ to achieve its URLLC requirement. In 5G standardization $L$ takes small values for reasonable signal strengths, i.e., for $\text{SNR}>0\text{dB}$ it is $L = 3$.}
Using \eqref{eq:success}, $R_t(\pi)$ can be computed as follows:
\[
R_t(\pi)=\mathbbm{1}\left\{\sum_{n=1}^N s_{k,n}(\pi)\geq L\right\}, ~~\forall \pi\in\mathcal{X}.
\]

To increase URLLC reliability, transmitter $k$ wants to maximize \emph{URLLC throughput} $\frac{1}T\sum_{t=1}^TR_t(x_k(t))$, where $T$ is some large integer that represents the horizon of interest for the application. 

\textbf{Energy.} We assume that the consumed energy is proportional to the rate of transmissions per frame, given by $\frac{1}T\sum_{t=1}^T\sum_{n=1}^N x_{k,n}(t)$.

In summary, the \emph{instantaneous utility} obtained by transmitter $k$ in frame $t$ is given by:
\begin{equation}\label{eq:obj}
U_{k,t}(x_k)=\underbrace{R_t(x_k(t))}_\text{URLLC thr.}-\eta_k \cdot \underbrace{\sum_{n=1}^N x_{k,n}(t)}_\text{energy cost},
\end{equation}
where scalar $\eta_k>0$ is a transmitter-selected weight that balances the importance of URLLC throughput and energy consumption. We mention that $U_{k,t}$ is unknown to transmitter $k$ since it depends on the patterns of all other users, via $R_k(x_k(t))$. 


\subsection{Problem formulation}

We would like to design a distributed protocol where each transmitter decides its pattern based only on the feedback of \textbf{(A.2)} in order to optimize the long-term average utility at some horizon $T$:
\begin{align*}
\displaystyle{\maximize_{x_k(1),\dots,x_k(T)\in \mathcal{X}^T} \frac{1}T\sum_{t=1}^T U_{k, t}(x_k(t))}
\end{align*}

 Random access protocols are expected to perform poorly w.r.t. this objective due to their following limitations. By design they do not ensure high latent throughput $R_t(x_k(t))$--as the number of transmitters increases the total latent throughput approaches zero--, they suffer from collisions and thus high energy levels per achieved throughput, and
 finally, they have limited flexibility and they are not adaptive to circumstances.
These considerations lead us to design a novel architecture, where each device performs an online learning algorithm in order to determine the most appropriate pattern for maximizing the obtained utility. 

\section{Architecture based on online learning}

We take the \emph{individual viewpoint of transmitter $k$} and optimize the utility $U_{k,t}(x_k(t))$ assuming that the rest transmitters are uncooperative, and their transmissions are seen as interference. In particular, to design an adaptive and robust algorithm, we will further assume that the other transmitters are adversaries that are choosing their patterns in order to lower $U_{k,t}(x_k(t))$. This worst-case approach will allow us to design an algorithm that is sensitive to interference and quickly adapts to changes in the environment.

\subsection{Restricting the design space}\label{sec:restr}

As in most learning problems, restricting the dimensions is essential for constructing an efficient  solution. In our problem, the number of possible patterns for transmitter $k$ is equal to the number of all possible binary vectors of length $N$, i.e., equal to $2^N$. For  values encountered in practice (e.g.~$N=100$) this creates an enormous action space. 


We introduce the concept  \emph{dictionary of patterns}, i.e., a  preselected subset of patterns $\mathcal{D}_k=\{\pi^1,\dots,\pi^d\}\subset \mathcal{X}$ of cardinality $d\ll 2^N$, to which transmitter $k$ will be restricted. The dictionary of patterns mimics the idea of the codebook in communications, where a subset of codes is designed off-line, and at runtime the transmitter selects a code from the codebook. 

\subsubsection{Basic rules for creating  dictionaries}
We provide some practical directions into creating pattern dictionaries.

\begin{itemize}
\item The zero pattern $(0,0,\dots,0)$ should always be included in the dictionary, since on many occasions a good action for user $k$ will be to remain silent within a frame.
\item Non-zero patterns with $\sum_{n=1}^N \pi_{k,n}<L$ should not be used, since they can not guarantee a successful frame and they consume more energy than the zero pattern.
\item Patterns with different values $\sum_{n=1}^N \pi_{k,n}\geq L$ should be used to allow exploration of protocols with different levels of energy and redundancy of transmissions.
\item For purposes of learning acceleration, the cardinality of the dictionary $d$ should be kept small, e.g. $d\leq d_{\max}$.
\item To avoid excessive number of collisions, it is preferable if different transmitters have different dictionaries. This can be achieved by generating the dictionaries in a random manner. However, we mention that having the same dictionary allows transmitters to share learned models, therefore the best approach would be to use groups od pseudo-random transmission patterns.
\end{itemize}

\subsubsection{Pattern dictionary design}

It is interesting to formulate the dictionary design as an optimization problem. However, we mention a few caveats. First, the optimization depends on the protocol of transmitters other than $k$, therefore this problem makes sense mostly when the rest of the transmitters have fixed and known protocols. Second, this is a combinatorial problem with non-convex objective and large dimensions, therefore a highly non-trivial optimization to solve.

Instead, we will take a very simple approach which appears to work in practice. 
We propose to use a simple, \emph{randomized}, and \emph{fully distributed} dictionary design algorithm. In particular, transmitter $k$ chooses its dictionary $\mathcal{D}_k$ by (i) including the zero pattern, (ii) excluding every pattern with less than $L$ transmissions, (iii) and then choosing the remaining $d-1$ patterns at random. Specifically, fix $d$ to be a large value which, however, will not slow down our algorithmic computations. For instance, a typical value could be between $100$ and $1000$. Start with an empty dictionary, i.e.,  $\mathcal{D}_k=\emptyset$. Also, recall that $L$ is  determined by the URLLC application. Then repeat the following steps:

\noindent \textbf{Randomized Dictionary Algorithm:}
\begin{enumerate}
\item Initialize dictionary with the zero pattern, i.e. $\mathcal{D}_k = \{\mathbf{0}\}$.  
\item Choose a number $\ell$ uniformly at random in $\{L,  \dots, N\}$ (the number of transmitting slots in a pattern).
\item Choose a random binary vector $\pi$ with $\ell$ ones (i.e. with $\ell$ transmitting slots).
\item If $\pi\notin \mathcal{D}_k$, then add it to the dictionary $\mathcal{D}_k\leftarrow \mathcal{D}_k\cup \{\pi\}$.
\item If  $|\mathcal{D}_k|=d$ stop.
\end{enumerate}

In the remaining we will assume that the dictionary of our transmitter is chosen with the above algorithm, and remains fixed for the playout of our protocol.


\subsection{Learning the best pattern in the dictionary} 

Consider a probability distribution $p=(p^1,\dots,p^d)$, where $p^i$ is a quality metric of pattern $\pi^i\in \mathcal{D}_k$. Learning the quality of  patterns in the dictionary consists in estimating a ``good'' probability distribution $p^*$ that would maximize the expected instantaneous utility:
\[
\overline{U}_t(p) = \sum_i p^i U_t(\pi^i).
\]
However, a complication arising in this paper is that the precise form of the utility $U_t(\pi^i)$ depends on the transmissions of all other users, and therefore it is unknown to the decision maker.

We will take the standard approach in the literature of Online Learning \cite{Shalev}. The idea is to allow $p$ to evolve over time, and at each iteration, to update it in a direction that improves the observed utility from the previous frame. The idea is that the previous frame serves as a ``prediction'' of what will happen in the next frame.

 Here, because the constraint for $p$ has the form of a simplex (a constraint $\sum_ip^i=1$), it is favorable to use the exponentiated gradient, instead of the classical gradient, see \cite{exp_num}.
Therefore, our update mechanism is as follows:
\[
p^i(t)=\frac{p^i(t-1)e^{-\alpha v^i}}{\sum_{j=1}^dp^j(t-1)e^{-\alpha v^j}},
\]
where the vector $v=(v^1,\dots,v^d)$ is a subgradient of $\overline{U}_{t-1}(p)$ at $p(t-1)$, and $\alpha$ is the learning rate.
Notice that the subgradient $v$ at frame $t$ is computed based on feedback obtained from the previous frame $t-1$. Specifically, 
the subgradient element $v^i$ has a very intuitive explanation as it is equal to the marginal benefit we would have in our expected utility (in the previous frame) if we would increase the probability of selecting pattern $\pi^i$. More simply, recall that $R_{t}(\pi)=1$ means that pattern $\pi$ achieves the URLLC objective in frame $t$, then we have $\forall i$:
\begin{equation}\label{eq:subgrad}
v^i = \left\{\begin{array}{rl}
-\eta_k\sum_n \pi_n^i & \text{ if } R_{t-1}(\pi^i)=0,\\
1-\eta_k\sum_n \pi_n^i & \text{ if } R_{t-1}(\pi^i)=1.
\end{array}\right.
\end{equation}

The learning rate $\alpha$ can be controlled to tradeoff how quickly and how accurately we learn. A typical choice in Online Learning is to optimize $\alpha$ for the horizon $T$, in which case we should choose:
\[
\alpha = \frac{\sqrt{2}}{G\sqrt{T}},
\]
where $G$ is an upper bound for each subgradient element. Hence, $G=\max\{1,\eta N\}$. Alternatively, the learning rate can be chosen larger to accelerate convergence (but discount the accuracy of convergence), or smaller to extend the convergence beyond the horizon (but make it more accurate). 

Some remarks are in order:
\begin{itemize}
\item The above algorithm is a variation of the online gradient algorithm of Zinkevich \cite{zinkevich}. At each iteration, the utility  $\overline{U}_t(p(t))$ is considered unknown (due to random or strategic transmissions of the other transmitters), and it is predicted using
\[\overline{U}_{t-1}(p(t-1))+\nabla\overline{U}_{t-1}(p(t-1))^T\left( p(t)-p(t-1)\right)
,\]
 which can be computed using the obtained feedback.
\item Specifically, our algorithm belongs to the category of \emph{Online Mirror Descent} algorithms (see \cite{Belmega,Shalev,exp_num}), which use gradient exponentiation. Such algorithms achieve the optimal learning rate in geometries with simplex constraints (such as in our case), while they do not require projection.
\end{itemize}

A common metric used to  quantify the quality of a learning algorithm is its \emph{regret}, which is defined as 
\[
\text{Regret}(T)=\sum_{t=1}^T \overline{U}_t(p^*)-\sum_{t=1}^T \overline{U}_t(p(t)),
\]
where $p(t)$ is the distribution chosen by a candidate algorithm, and $p^*$ is the best distribution if we would know the entire sequence of transmissions  of all other transmitters over the entire horizon $T$. Standard results from the literature of online learning tell us that our algorithm minimizes the worst-case regret and achieves $\text{Regret}(T)=o(T)$, i.e., (1) our algorithm is the best learner in the case that the other transmitters are trying to hurt us, and (2) as frames evolve, we learn the best static distribution $p^*$.

At this point, we mention that although the other transmitters \emph{are not really manipulated by an adversary}, our algorithm is so sensitive to changes in the interference that it can optimally adapt to many different scenarios, and in particular to situations that the interference fluctuate in a very abrupt and non-stationary way.

\section{The Learn2MAC Access Protocol}

In this section we summarize the design of our online learning-based multiple access protocol. The procedure is shown as Algorithm 1. 

\begin{algorithm}[h!]
	\caption{Learn2MAC}\label{alg:Qlearning}
	\begin{algorithmic}[1]	
		\State {Choose a $d$ (typically as large as possible while the algorithm runs efficiently).}
		\State {Choose the dictionary $\mathcal{D}_k\subseteq \mathcal{X}$ with $|\mathcal{D}_k|=d$ using the ``randomized dictionary algorithm'' above.}
		\State {Initialize $\alpha=\sqrt{2/(T\max\{1,\eta^2N^2\})}$, $p_k(0)=(\frac{1}d,\dots,\frac{1}d)$.}
		\For {every frame $t=1,\dots,T$} 
			\State {Update the probability distribution $p_k(t)$ using:
\[
p^i_k(t)=\frac{p^i_k(t-1)e^{-\alpha v^i_k}}{\sum_{j=1}^dp^j_k(t-1)e^{-\alpha v^j_k}},\quad i=1,\dots,d.
\]
						 }
			\State {Choose a pattern from $\mathcal{D}_k$ at random according to the distribution $p_k(t)$.}
			\State {Transmit according to the chosen pattern.} 
		\EndFor 	
	\end{algorithmic} 
\end{algorithm}

Above, we use the following notation:
\begin{itemize}
\item $\mathcal{D}_k$ is the dictionary of patterns, see Sec.~\ref{sec:restr},
\item $d$ is the size of the dictionary. 
\item $\alpha$ is the learning rate,
\item $p_k(t)$ is a probability distribution over the patterns of the dictionary, and
\item $v_k$ is the subgradient vector in frame $t$, see \eqref{eq:subgrad}.
\end{itemize}

As a final remark, note that Learn2MAC exploits the fact that the feedback received is the occupancy of the medium at each slot within the frame, therefore can be used to deduce the performance of \emph{every} transmission pattern (and not the one just used) in the previous frame. This helps significantly speed up the learning process, and therefore the adaptability of the algorithm in changing environments. 

\section{Numerical Analysis}

In this Section we illustrate the performance of Learn2MAC and its superiority with respect to baseline random access schemes via simulations. All simulations lasted for $T=30000$ frames. The setting here is that each frame has length of $N = 20$ slots, a URLLC packet of device $k$ is  delivered if at least $L=2$ transmissions in the frame were successful, and a device using Learn2MAC has a dictionary of $d=100$ transmission patterns. The weight balancing the importance or latent throughput vs. energy consumption is set to $\eta_k=0.05$ for each device. Finally, the learning rate is set independently of the simulation horizon (which is quite relevant in practice since it may not be easy/possible to know how many frames a user will be active i advance) to $\alpha=0.001$. We compare Learn2MAC vs. the use of a standard random access scheme, where the device transmits at each slot independently at random with a probability $\overline{q}$.

We first verify that a single device using Learn2MAC can adapt to an environment with devices using a pre-existing protocol. For this, we examine two cases: (i)"Static Interference", where half of the slots of a frame are pre-allocated  in a fixed TDMA fashion, and (ii) "Dynamic Interference" where pre-existing terminals access each slot of the frame randomly, each with a probability that is periodic in time. For a fair comparison, the access probability $\overline{q}$ of the baseline random access scheme is configured so that the energy expenditure is the same in both cases.  

Results on the running average URLLC throughput are shown in Figures 2 and 3, respectively. Regarding the first case, Fig. 2 illustrates clearly that Learn2MAC learns to use the pattern which corresponds to transmissions in slots left idle by the background TDMA schedule; we also observed  that, moreover, Learn2MAC learns the most efficient such code (i.e. the one with $2$ transmissions in idle slots). By contrast, the random access baseline performs very poorly. Regarding the case with dynamic background user activity, Fig. 3 illustrates that Learn2MAC achieves a higher throughput than the random access baseline for the same energy expenditure, therefore adapting transmissions to economically use energy in this case as well.   

We then compared the two protocols for the case of uncoordinated medium access; herein, we have $K$ devices using Learn2MAC in one case and a random access protocol with transmission probability $\overline{q}=0.2$ at each slot \footnote{This value was chosen because it provides a good balance between  not even attempting to transmit at least $L=2$ times (due to access probability being too low), thus losing latent throughput, and transmitting too aggressively, thus leading to many collisions.} in the other. We run the simulations for $T=30000$ frames as above and measure the total URLLC throughput obtained by the system (by summing up the URLLC throughput obtained by each device) at the end of each run for different number of devices $K$. These results are shown in Fig. 4. Remark that, since $L=2$ and there are $N=20$ slots in each frame, the maximun number of devices (scheduled by a centralized controller in non-overlapping slots) successfully transmitting a URLLC packet is $10$ per frame, which is the upper bound on the total latent throughput. From Fig. 4 we can observe that, at relatively low and medium load (up to $K=7$ devices), the total latent throughput scales almost \emph{linearly} with $K$: this means that Learn2MAC enables the devices to learn to use transmission patterns with no or little overlap with respect to each other, thus had few collisions and the devices were able to all coexist and transmit their URLLC packets in almost every frame. By contrast, when random access is used, the throughput obtained is still low due to collisions. When the number of devices approaches $10$, which is the maximum that can be supported, Learn2MAC exhibits the classical behaviour of uncoordinated medium access algorithms - namely rapid decrease in the latent throughput of the system due to collisions (while still outperforming the random access baseline though). This is the regime where admission control is really needed, since the available resources are very close to (or less) than the total needed by the devices and Learn2MAC still leads to many collisions in this case. This result suggests that Learn2MAC should be augmented by a mechanism where devices learn if the system is in the high- or low- load regime, and some devices must learn to completely disconnect from the system if the former is the case. This direction is very interesting from both the algorithmic/theoretical and the practical perspective and we leave it as future work.            

\section{Conclusion}

\begin{figure}
	\centering 
	\includegraphics[scale = 0.25]{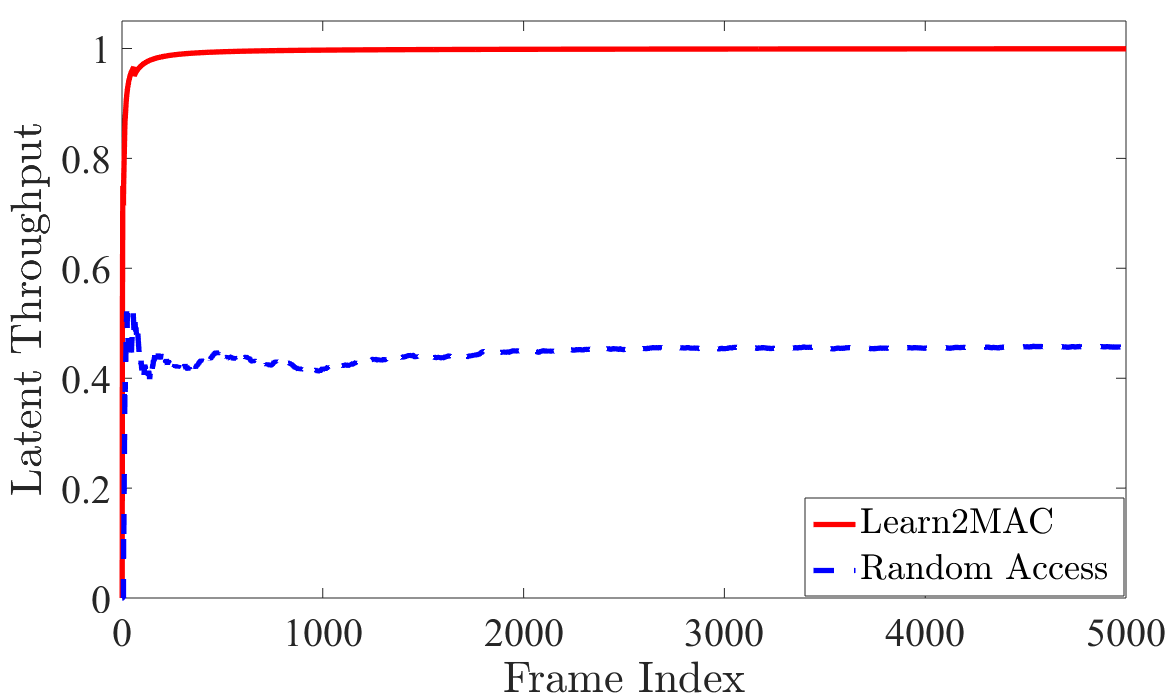}
	\caption*{\small {\bfseries Fig. 2 (Latent Throughput under TDMA Interference):} Running average of the URLLC throughput obtained from a single device using Learn2MAC and a baseline (ALOHA) random access scheme with a TDMA background schedule. The access probability of the baseline scheme is such that it results to the same energy expenditure as Learn2MAC.} \label{fig:vsTDMA}	
\end{figure}

\begin{figure}
	\centering 
	\includegraphics[scale = 0.25]{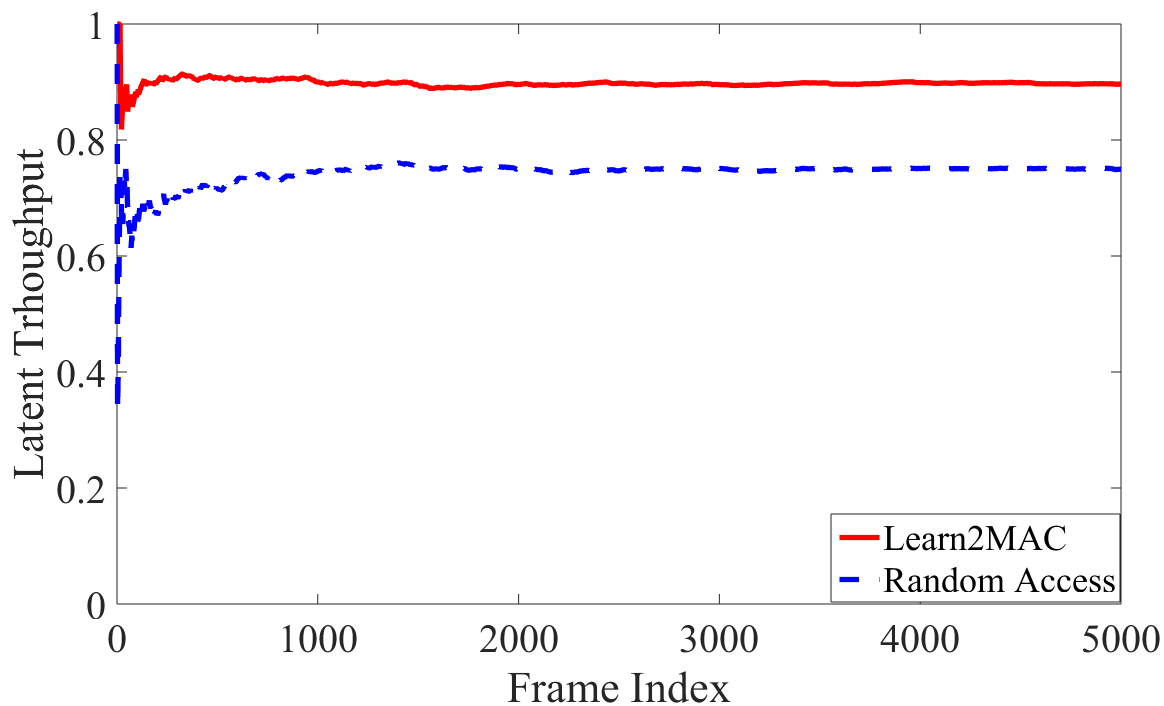}
	\caption*{\small {\bfseries Fig. 3 (Latent Throughput under ALOHA Interference):} Running average of the URLLC throughput obtained from a single device using Learn2MAC and a baseline (ALOHA) random access scheme with a background ALOHA scheme with periodic access probabilities.The access probability of the baseline scheme is such that it results to the same energy expenditure as Learn2MAC.} \label{fig:vsDynamic}	
\end{figure}

\begin{figure}
	\centering 
	\includegraphics[scale = 0.25]{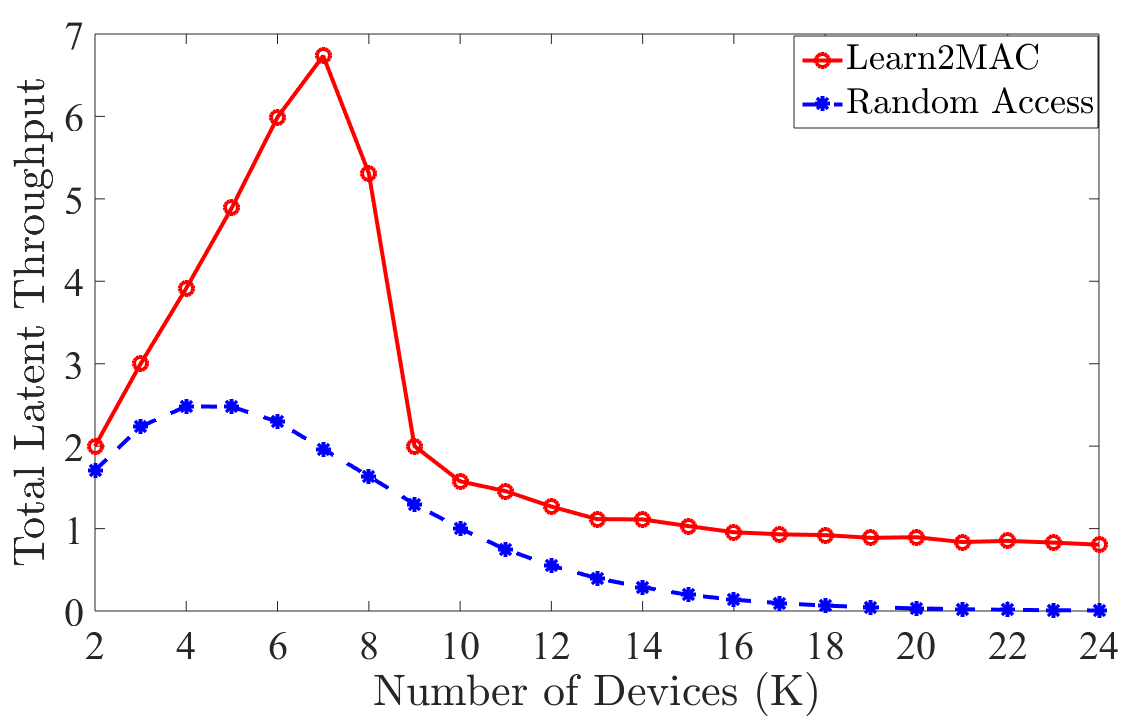}
	\caption*{\small {\bfseries Fig. 4 (Saturation Latent Throughput Analysis):} Comparison of the system's performance between the cases where  (i) all devices user Learn2MAC and (ii) all devices use a baseline (ALOHA) random access protocol with transmission probability $\overline{q}=0.2$. The maximum number of users that can be scheduled in a way that achieves their URLLC transmission requirement in this simulation setting is $10$.}  
	\label{fig:satPlot}		
\end{figure}

In this paper, we proposed Learn2MAC, an Online Learning-based Multiple Access schemes that allows users to decide in a distributed manner which transmission pattern to choose. It is shown that Learn2MAC can provide URLLC guarantees, which is an important limitation of other uncoordinated access schemes, and outperform standard random access both in cases where a single device needs to adapt, in an energy-efficient manner, to an environment with users following pre-existing and in cases where multiple devices need to coordinate using the same protocol. In the latter case, it can enable devices to learn to coordinate with almost $100\%$ latent throughput in cases with high and medium number of resources. Therefore, Learn2MAC is a strong candidate for IoT applications that require at the same time latency guarantees, energy efficiency, and low coordination overhead.  


\bibliography{AIMA_biblio}

\begin{thebibliography}{10}
\providecommand{\url}[1]{#1}
\csname url@samestyle\endcsname
\providecommand{\newblock}{\relax}
\providecommand{\bibinfo}[2]{#2}
\providecommand{\BIBentrySTDinterwordspacing}{\spaceskip=0pt\relax}
\providecommand{\BIBentryALTinterwordstretchfactor}{4}
\providecommand{\BIBentryALTinterwordspacing}{\spaceskip=\fontdimen2\font plus
\BIBentryALTinterwordstretchfactor\fontdimen3\font minus
  \fontdimen4\font\relax}
\providecommand{\BIBforeignlanguage}[2]{{%
\expandafter\ifx\csname l@#1\endcsname\relax
\typeout{** WARNING: IEEEtran.bst: No hyphenation pattern has been}%
\typeout{** loaded for the language `#1'. Using the pattern for}%
\typeout{** the default language instead.}%
\else
\language=\csname l@#1\endcsname
\fi
#2}}
\providecommand{\BIBdecl}{\relax}
\BIBdecl

\bibitem{report}
\BIBentryALTinterwordspacing
``Ericsson mobility report,'' June 2018. [Online]. Available:
  \url{https://www.ericsson.com/en/mobility-report}
\BIBentrySTDinterwordspacing

\bibitem{pure_aloha}
N.~Abramson, ``The {ALOHA} system: Another alternative for computer
  communications,'' in \emph{Proceedings of the November 17-19, 1970, Fall
  Joint Computer Conference}, ser. AFIPS '70 (Fall), 1970, pp. 281--285.

\bibitem{aloha_slotted}
L.~G. Roberts, ``{ALOHA Packet System with and Without Slots and Capture},''
  \emph{SIGCOMM Computer Communications Review}, vol.~5, no.~2, pp. 28--42,
  Apr. 1975.

\bibitem{bianchi00}
G.~Bianchi, ``Performance analysis of the {IEEE} 802.11 distributed
  coordination function,'' \emph{IEEE Journal on Selected Areas in
  Communications}, vol.~18, no.~3, pp. 535--547, March 2000.

\bibitem{NiSrikant}
J.~Ni, B.~Tan, and R.~Srikant, ``{Q-CSMA: Queue-Length-Based CSMA/CA Algorithms
  for Achieving Maximum Throughput and Low Delay in Wireless Networks},''
  \emph{IEEE/ACM Transactions on Networking}, vol.~20, no.~3, pp. 825--836,
  June 2012.

\bibitem{JiangWalrand}
L.~Jiang and J.~Walrand, ``{A Distributed CSMA Algorithm for Throughput and
  Utility Maximization in Wireless Networks},'' \emph{{IEEE/ACM} Transactions
  on Networking}, vol.~18, no.~3, pp. 960--972, Jun. 2010.

\bibitem{SuccesiveCanc}
F.~Clazzer, E.~Paolini, I.~Mambelli, and {\v{C}}.~Stefanović, ``Irregular
  repetition slotted {ALOHA} over the {Rayleigh} block fading channel with
  capture,'' in \emph{2017 IEEE International Conference on Communications
  (ICC)}, May 2017.

\bibitem{Apostolos2}
A.~Destounis and G.~S. Paschos, ``{Complexity of URLLC Scheduling and Efficient
  Approximation Schemes},'' in \emph{International Symposium on Modeling and
  Optimization in Mobile, Ad Hoc, and Wireless Networks (WiOpt)}, submitted.

\bibitem{Apostolos1}
A.~Destounis, G.~S. Paschos, J.~Arnau, and M.~Kountouris, ``Scheduling {URLLC}
  users with reliable latency guarantees,'' in \emph{International Symposium on
  Modeling and Optimization in Mobile, Ad Hoc, and Wireless Networks (WiOpt)},
  May 2018.

\bibitem{Shalev}
S.~Shalev-Shwartz, ``{O}nline {L}earning and {O}nline {C}onvex
  {O}ptimization,'' \emph{Foundations and Trends{\textregistered} in Machine
  Learning}, vol.~4, no.~2, pp. 107--194, 2012.

\bibitem{exp_num}
L.~Vigneri, G.~Paschos, and P.~Mertikopoulos, ``{Large-Scale Network Utility
  Maximization: Countering Exponential Growth with Exponentiated Gradients},''
  in \emph{IEEE International Conference on Computer Communications (INFOCOM)},
  May 2019.

\bibitem{zinkevich}
M.~Zinkevich, ``{O}nline {C}onvex {P}rogramming and {G}eneralized
  {I}nfinitesimal {G}radient {A}scent,'' in \emph{International Conference on
  International Conference on Machine Learning (ICML)}, 2003.

\bibitem{Belmega}
\BIBentryALTinterwordspacing
E.~V. Belmega, P.~Mertikopoulos, R.~Negrel, and L.~Sanguinetti, ``{O}nline
  {C}onvex {O}ptimization and {N}o-{R}egret {L}earning: {A}lgorithms,
  {G}uarantees and {A}pplications,'' \emph{CoRR}, vol. abs/1804.04529, 2018.
  [Online]. Available: \url{http://arxiv.org/abs/1804.04529}
\BIBentrySTDinterwordspacing

\end{thebibliography}
\bibliographystyle{IEEEtran}
\end{document}